\documentstyle[11pt]{article}
\begin{document}

\def\msbar{{\rm \overline{MS\kern-0.14em}\kern0.14em}}

\begin{flushright}
 HEP-LAT Preprint: hep-lat/9209008\\
 Liverpool Preprint: LTH 285\\
 Edinburgh Preprint: 92/512\\
August 27, 1992

\end{flushright}
\vspace{5mm}
\begin{center}

{\LARGE\bf The Running Coupling from SU(3) Lattice Gauge Theory}\\[1cm]

{\large\it UKQCD Collaboration}\\[3mm]

{\bf S.P.~Booth,  D.S.~Henty }\\
Department of Physics, The University of Edinburgh, Edinburgh EH9~3JZ,
Scotland

{\bf A.~Hulsebos, A.C.~Irving, C.~Michael, P.W.~Stephenson}\\
DAMTP, University of Liverpool, Liverpool L69~3BX, UK

\end{center}

\begin{abstract}

 From an accurate determination of the inter-quark potential, one can
study the running coupling constant for a range of $R$-values
and hence estimate the scale $\Lambda_{\msbar} $.
Detailed results are  presented  for $SU(3)$  pure  gauge  theory.

\end{abstract}

\section{Introduction}
\par
In the continuum the potential between static quarks is known
perturbatively to two loops in terms of  the  scale $\Lambda_{\msbar} $.
 For  $SU(3)$ colour, the continuum force is given by \cite{bill}
$$
{dV \over dR } =  {4 \over 3} {\alpha(R) \over R^2}
$$
\noindent with the effective coupling $\alpha (R)$ defined as
$$
\alpha (R) = { 1 \over 4 \pi [ b_0  \log (R\Lambda _R )^{-2} +
(b_1 / b_0 ) \log \log (R\Lambda_R )^{-2} ] }
$$
\noindent where $b_0=11/16 \pi ^2$ and $b_1=102 \ b_0^2/121$ are
 the usual coefficients in the
perturbative expression for the $\beta$-function
and, neglecting quark loops in the vacuum,  $\Lambda_R= 1.048
\Lambda_{\msbar}$.
Note that the usual lattice regularisation scale $\Lambda_L = 0.03471
 \Lambda_{\msbar}$.
\par
 At large separation $R$, the potential behaves as $KR$ where $K$  is  the
string tension.  Thus in principle knowledge of  the  potential $V(R)$
 serves  to
determine the dimensionless ratio $\sqrt K/\Lambda $ which relates the
perturbative scale $\Lambda$ to a non-perturbative observable such as the
string tension $K$.  This is the  basis  of
the method~\cite{cmlms} used for $SU(2)$ which we extend here to $SU(3)$
colour.

For $SU(3)$ as for $SU(2)$, the bare lattice coupling proves to be a
poor guide to physical behaviour in that asymptotic scaling to two loops
is not yet manifest. However, the weaker requirement of scaling is well
satisfied:  the dimensionless ratios of physical quantities are
found to be independent of $\beta $.  For example in $SU(2)$, the
potential $V(R)$ scales
\cite{ukqcd} over a range of lattice spacing of a factor of 4 (from
$\beta =2.4$ to $ 2.85$).  That scaling but not asymptotic scaling is
valid implies that the bare coupling constant derived from $\beta $ is
inappropriate and that an effective coupling constant derived from some
physical quantity is a better choice.  This has been emphasized by
Lepage and Mackenzie \cite{lm}.  It is also the basis of the method
proposed by L\"uscher et al.~\cite{lu} to extract the running coupling
constant.

Here we choose to determine the running coupling constant from the
interquark potential between static quarks at small distance $R$.  This
quantity can be determined in a straightforward way from lattice
simulation on large volume lattices.  Although we require small $R$ and hence
large energy $1/R$ to make most precise contact with the perturbative
expression, the lattice method implies the presence of lattice
artefacts when $R \approx a$. Thus we need to work on the largest
spatial lattice available, consistent with avoiding finite size effects.
We present results from a $36^4$ lattice at $\beta=6.5$ to achieve
this. These results are compared with previous UKQCD data~\cite{uksu3}
from $24^3\times 48$ lattices at $\beta=6.2$ to check scaling.

\section{Lattice potentials}

Our main aim is to explore the interquark potential at as large a
$\beta$-value as feasible. We wish to remain in the large volume region
so memory constraints are limiting. On our Meiko Computing Surface with
64 i860 processors, each having 16 megabytes main memory, we are able to
update at most a $36^4$ $SU(3)$ lattice efficiently.  Previous work
\cite{su3,uksu3,bali} has shown that at $\beta =6.5$ such a lattice
should be suitable.

The difficult part of the analysis is to determine the string tension
accurately on such lattice configurations. In order to measure the
largest number of configurations within our data communication
constraints, we use rather similar methods to those we used to determine
the string tension in $SU(2)$~\cite{ukqcd}. We first combine the spatial
links of the lattice once~\cite{teper} using a sum of straight and
U-bends of length 2 to an effective $18^3 \times 36$ lattice.  The
spatial links of these lattices are then further smeared repeatedly (APE
smearing \cite{ape} with $SU(3)$ projection of $2.5
\times$ straight link plus four spatial U-bends) to build up paths
between the static sources.  Building up spatial paths from these links,
we measure generalised Wilson loops to determine the potentials at
on-axis separations with $R/a$ = 2, 4, 6, 8, 10, 12, 14, 16, 18, 20, 22, 24.
We also measure at planar off-axis separations with $R/a$-vectors
(2,2,0), (4,2,0), (4,4,0), (6,2,0), (6,4,0) and (6,6,0).  We use a
recursive smearing with 40 and 10 levels --- this provides a $2 \times 2$
basis for the standard variational technique of finding the optimum
combination of these two spatial paths at each $R$-value.

We measure potentials for the $R$-separations listed above and
$T$-separations 0 to 7.  The lattice is well equilibrated with 1200
heat-bath sweeps and configurations are measured every 150 sweeps
thereafter, using 4 pseudo-overrelaxation sweeps for every heatbath
sweep.  `Pseudo-overrelaxation' refers to performing $SU(2)$
overrelaxation steps in 3 subgroups of $SU(3)$. We are able to perform a
complete updating step in 27 seconds for pseudo-relaxation and in 30
seconds for the heatbath.  The similarity between these timings may seem
surprising, but one major reason is that message-passing between
processors is the most significant factor in both cases.  All the
CPU-intensive parts of our programme are written in well-optimised
assembler code. We obtain an average action $S=0.63835(2)$.

We measure the on-axis potentials on 50 configurations. With such
limited statistics an accurate determination of the autocorrelation
is not possible, but our measurements are consistent for all observables
 with an
autocorrelation time of less than 225 sweeps.  Our error analysis
 is obtained from bootstrap analysis of these 50 measurements which
we treat as being statistically independent since we
find equivalent results when using 25 blocks of 2 measurements.

  The
potentials are given by the extrapolation in $T$ of the ratio of
generalised Wilson loops.
 $$
V(R) = \lim_{T \to \infty} V(R,T),
 $$
where
 $$
V(R,T)= -\log W(R,T)/W(R,T-a)
 $$
We take the optimum linear combination of paths to determine the ground
state potential where we use the 2/1 $T$-ratio for this purpose since
the statistical errors are quite small at those $T$-values. We find that
this path combination gives an overlap of $85\%$ or more.  Now there is
a monotonic decrease with $T$ of the effective potential $V(R,T)$
evaluated from the correlations ($W(R,T)$) in that path eigenvector.
One way to select the limit at large $T$ is to identify a plateau where
$V(R,T)$ and $V(R,T+a)$ analyses give consistent values within the error
on the difference. We find that the 4/3 $T$-ratio is consistent with the
start of such a plateau.  As a check on possible systematic errors
arising from this $T$-extrapolation we can try to fit the decrease of
$V(R,T)$ with $T$ using the energy gap to the first excited state.  For
the larger $R$ potentials where this extrapolation in $T$ may introduce
sizeable errors, we obtain estimates of this energy gap from our
variational method in the 2:1 $T$-ratio basis and these estimates agree
with the string model estimate $2\pi/R$.  We then use those estimates to
obtain a systematic error associated with the $T$-extrapolation.  We
find consistency between the effective potential based on $T$-values 2-4
and on $T$-values 3-5 which confirms the stability of the method.  In
order to represent the data in table form it is better to give the force
since the errors there are less correlated between different $R$-values.
The force derived from our potential measurements of adjacent
$R$-values with $T$-ratio 4/3 is
shown in table~\ref{force}.

We fit the $R$-dependence of the potential at large $R$ to obtain the
string tension $K$ with
$$
  V(R) = C - { E \over R } + K R
$$
where for the Coulomb component we use a lattice one gluon exchange
propagator (see below) even though at large $R$ this is very close
to the continuum expression $1/R$.
When we take full account in the fit of the statistical
correlations between Wilson loops at different $R$-values, we get an
acceptable $\chi^2$ but some sign of instability in inverting the
$15 \times 15$ correlation matrix. Fitting instead to the force
with diagonal errors gives similar results which are more stable.
These results to a fit for $R \ge 4a$ are shown in table~\ref{parek}.

As an estimate of the systematic error coming from $T$-extrapolation,
we fitted potential values from several different prescriptions. Using
$T$-ratio 5/4 gave $Ka^2=0.0110(3)$; extrapolating
(as discussed above) the $T$ 2-4 potentials gave 0.0111(4) while
extrapolating 3-5 potentials gave 0.0107(8). Hence the systematic
error from $T$-extrapolation appears comparable to the statistical
error and we quote $Ka^2=0.0114(4)$ as an overall determination.

\par
\begin{table}
\centering
\begin{tabular}{|r|l|}\hline
 $R/a$ \  & $\ \Delta V/\Delta R $    \\\hline
 2.4142 & 0.0667(4)   \\
 3.4142 & 0.0344(4)   \\
 4.2361 & 0.0286(8)   \\
 5.0645 & 0.0226(5)  \\
 5.8284 & 0.0196(19)  \\
 6.1623 & 0.0190(16)  \\
 6.7678 & 0.0174(9)   \\
 7.6056 & 0.0161(10)  \\
 8.2426 & 0.0161(11)  \\
 9.0000 & 0.0149(3)   \\
11.0000 & 0.0130(4)   \\
13.0000 & 0.0134(5)   \\
15.0000 & 0.0121(5)   \\
17.0000 & 0.0132(6)   \\
19.0000 & 0.0117(6)   \\
21.0000 & 0.0125(8)   \\
23.0000 & 0.0126(7)     \\\hline
\end{tabular}
\caption{ \label{force}
 The force $\Delta V/\Delta R$ at
average separation $R$ derived from $T$-ratio 4/3.
}
\end{table}

\begin{table}
\centering
\begin{tabular}{|c|c|c|}\hline
      $E$      &      $Ka^2$ & $\chi^2 / {\rm d.o.f.}$\\\hline
    0.278(7)  &     0.0114(2)&   9.1/13 \\ \hline
\end{tabular}
\caption{ \label{parek}
The fitted parameters to the potential for $R\ge 4a$.
}
\end{table}

\par
\begin{table}
\centering
\begin{tabular}{|r|l|c|l|}\hline
 $R/a$ \  & $\ \Delta V/\Delta R $  & $\Delta V_c / \Delta R$ &
 $ \qquad \alpha(R)$  \\\hline
 1.2071 & 0.2067(7)    & 0.1607    & 0.170(1)(5)    \\
 1.7071 & 0.0750(7)    & 0.0930    & 0.197(1)(4)    \\
 2.1180 & 0.0959(19)   & 0.0664    & 0.223(6)(10)   \\
 2.5322 & 0.0541(5)    & 0.0523    & 0.248(2)(1)   \\
 2.9142 & 0.0263(40)   & 0.0424    & 0.270(26)(10)  \\
 3.0811 & 0.0471(39)   & 0.0368    & 0.262(28)(7)  \\
 3.3839 & 0.0391(12)   & 0.0371    & 0.317(10)(2)   \\
 3.9241 & 0.0292(5)    & 0.0290    & 0.333(6)(1)  \\
 \hline
\end{tabular}
\caption{ \label{alphat}
The force $\Delta V/\Delta R$ and lattice artefact corrected force
$ \Delta V_c / \Delta R $ at
average separation $R$. The running coupling $\alpha (R)$ derived
from the corrected force is shown as well. The second error shown on
$\alpha$ is $10\%$ of the lattice artefact correction.
}
\end{table}

\begin{table}
\centering
\begin{tabular}{|c|c|c|c|}\hline
$A$ & $f$ & $Ka^2$ & $B/a$ \\\hline
 0.311(14) & 0.64(6) & 0.01138 & 0.067(13) \\\hline
\end{tabular}
\caption{ \label{parf}
Fit to force from table~\protect\ref{alphat} for $R>a$.
}
\end{table}

As well as such a high statistics study of the potential at large $R$,
we determine the small $R$ potential at a variety of on- and off-axis
$R$-values. These have separations with $R/a$ vectors (1,0,0), (1,1,0),
(2,0,0), (2,1,0), (2,2,0), (3,0,0),(3,1,0), (3,2,0) and (3,3,0).  For
this analysis we use the full spatial configuration and a recursive
smearing~\cite{ape} with $c=2.0$ and 30 iterations.  Since this small
$R$-potential is easily measured, we use lower statistics. We evaluate 4
configurations (as 8 half-configurations for error analysis).  The force
obtained from these results using $T$-ratio 4/3 is shown in
table~\ref{alphat}. Results for the potential at common $R$-values
between the two methods are consistent, with the latter method giving a
larger overlap (greater than $97\%$).

The potential shows a lack of rotational invariance at small $R$.
To lowest order this can be attributed to the difference $\delta G(R)$ between
the lattice one gluon exchange expression and the continuum expression.

$$
\delta G(R) =  {4 \pi \over a} \int_{-\pi}^{\pi}
 { d^3 k \over(2\pi)^3}
{ e^{ik.R/a} \over 4\sum_{i=1}^3 \sin^2 (k_i / 2) }
-{1 \over R}
$$
On a lattice, the next order of perturbation has been calculated \cite{hk}
and the dominant effect is a change from the bare coupling
to an effective coupling \cite{lm}. In that case,  using the difference
above but with an adjustable strength will correct for the small $R/a$
lack of rotational invariance. A test of this will be that a smooth
interpolation of $V(R)$ versus $R$ is obtained with this one free
parameter to the 6 off-axis potential values.

We  evaluate $\delta G(R)$
numerically using the limit of a very large lattice since we are not
here concerned with long-range effects.  Fixing the string tension $K$
 at the value found in the large $R$ fit,  we find the  following
empirical expression provides a good fit to the data of
table~\ref{alphat} for $R~> a$,
$$
aV(R)=C- {A \over R}+{B \over R^2}+KR - A f \delta G(R),
$$
with $\chi ^{2}$ per degree of freedom 4.1/4. The fit parameters are
shown in table~\ref{parf}. For our present purposes the detailed form of
this fit at small $R$ is not relevant --- what is needed is a
confirmation that a good fit can be obtained. This then supports our
prescription to correct the lattice artefacts responsible for the lack
of rotational invariance. What is more difficult is to assign errors to
this correction procedure. We follow the $SU(2)$ analysis \cite{ukqcd}
in using as an illustration a $10\%$ systematic error on the artefact
correction itself with the proviso that for the lowest $R$ value ($R=a$)
the smooth interpolation is less of a constraint so that we disregard
that datum in the analysis.  We then assume that an improved estimate
for the continuum potential $V_c$ will be obtained by correcting the
measured lattice values $V$ by $\delta G$ with the fitted coefficient.
These values are shown in table~\ref{alphat}.

\section{Running Coupling}

It is now  straightforward  to  extract  the  running  coupling
constant by using
$$
\alpha( { R_1 + R_2 \over 2 }) = { 3\over 4} R_1 R_2 { V_c(R_1)-V_c(R_2) \over
 R_1-R_2 }
$$
\noindent where the error in using a finite difference is here negligible.
This is shown in table~\ref{alphat} and is plotted  in the figure versus
$R \sqrt K$ where $K$ is taken from the fit - see table ~\ref{parek}.
The interpretation of $\alpha$ as defined above as an
effective running coupling constant is only justified at small $R$ where the
perturbative expression dominates.
  Also shown  are  the
two-loop perturbative results for $\alpha(R)$ for
different values of $\Lambda_R $.

\begin{figure}
\centering
\vspace{10cm}
\includegraphics{figsu3.ps}
\caption{
The effective running coupling constant $\alpha(R)$ obtained from
the force between static quarks at separation $R$.  The scale is set
 by the string tension $K$. Data at $\beta=6.5$ are from
table~\protect\ref{alphat} (diamonds) and at $\beta=6.2$ (triangles).
 The dotted error bars represent an estimate
of the systematic error due to lattice artefact correction as described
in the text.  The curves are the two-loop perturbative expression with
$a(6.5)\Lambda_R=0.060$ (dotted) and 0.070 (continuous).
}
\end{figure}

The figure clearly shows a {\it running} coupling constant.  Moreover
the result is consistent with the expected perturbative dependence on
$R$ at small $R$.  There are systematic errors, however. At larger $R$,
the perturbative two-loop expression will not be an accurate estimate of
the measured potentials, while at smaller $R$, the lattice artefact
corrections are relatively big.  Setting the scale using $\sqrt K=0.44$
GeV implies $1/a(\beta=6.5)=4.13 $ GeV, so $R < 4a(6.5)$ corresponds to values
of $1/R > 1$ GeV.  This $R$-region is expected to be adequately
described by perturbation theory.  Another indication that perturbation
theory is accurate at such $R$-values is that $\Delta V_c / \Delta R$ at
small $R$ is found to be very much greater than the non-perturbative
value $K$ at large $R$.

Even though the lattice artefact correction of all 6 off-axis points by
one parameter is very encouraging, the only way to be certain that
lattice artefacts are eliminated is by the comparison of different
$R/a$ values at the same physical
$R$ value and this can be  achieved by using different $\beta$ values.
 Now this test was satisfied in an $SU(2)$
study~\cite{cmlms,ukqcd}.  Even so we can check independently in $SU(3)$
and we use UKQCD data \cite{uksu3} at $\beta =6.2$. This data comes from
measuring 30 well separated $24^3 \times 48$ lattice configurations with
smearing~\cite{ape} with $c=4.0$ and 28 and 40 recursive iterations for
a 2 path basis. Both on- and off-axis potentials are evaluated.  We
fitted the potentials from $a < R \le 12a $ with a 4-parameter
expression to take account of the lattice artefacts. From $T$-ratio 4/3
we get $Ka^2 = 0.0251(5)$, while the 5/4 $T$-ratio yields 0.0252(8). The
3/2 and 4/3 $T$ extrapolation method gave $Ka^2=0.0239(11)$. These
results are consistent and we use a compromise value of 0.0251(8). This
corresponds to a ratio of lattice spacings $a(6.2)/a(6.5)=1.484(35)$ to
be compared with the two-loop perturbative ratio of 1.404.  Setting the
scale from the measured string tensions, we also show the $\beta =6.2$
results for the effective running coupling in the figure. There is
excellent agreement with the results from $\beta=6.5$.

The easiest way to describe the value of the running coupling constant
$\alpha$
is in terms of a scale or $\Lambda$ value with the understanding that
we are only determining $\alpha$ for a range of energy scales
 $1/R$ - namely 1 to 3 GeV.
The final estimate of $\Lambda$ is made from the figure, weighting
smaller $R$ more heavily since the perturbative expression is
more accurate as $\alpha(R)$ becomes smaller. We exclude the lowest
$R$ point since the lattice artefact correction is for $R>a$.
  Remembering that the systematic errors due
to lattice artefact correction are estimates only and since these systematic
errors are dominant, we do not attempt a fit but we can  conclude
that our results are consistent with values of
 $\Lambda$ lying in the range shown by the two curves plotted.
{}From the data at $\beta $ = 6.5, these curves have
$a(6.5)\Lambda_R$=0.070 and 0.060.  Using the value of the string
tension from the fit, we get $\sqrt K/\Lambda_L$= 49.6(3.8).  Moreover, this
value is consistent with the evaluation at both $\beta=$ 6.5 and 6.2.

\section{Conclusions}

Using  the bare
coupling $g$ derived from $\beta =6/g^{2}$ and the  two-loop  perturbative
 relationship  $a(g)$ in terms of the scale $\Lambda_L$
gave \cite{su3,uksu3,bali} the following
slowly decreasing values of $\sqrt K/\Lambda_L =$ 93.0(7) and
96.7(1.6)(2.6) at $\beta=6.0$; 85.9(1.5) and 86.4(1.0)(1.9) at $\beta=
6.2$ and
82.3(8)(1.7)
 at $\beta = 6.4$.  Our present analysis
at $\beta=6.5$ yields $\sqrt K/\Lambda_L =80.0(1.4)$. Clearly, the
$\beta \rightarrow \infty $ limit lies below these values. Moreover
the statistically significant decrease of these values is evidence that
two-loop perturbative scaling is not obtained in terms of the bare coupling.
Our present method which does not rely on the bare coupling
gives  the scaling result which should be independent of $\beta$.
Our estimate is  $ \sqrt K /\Lambda_L = 49.6(3.8)$.    This
is sufficiently far below the values extracted from the bare coupling
to imply  that asymptotic scaling to two-loop perturbation theory is
 not ``just around the corner'' but will only  be
satisfied accurately at larger $\beta$-values than those currently
accessible to lattice simulation.

Our result for the running coupling $\alpha_R(R)$ given in the figure and
table~\ref{alphat} can be read directly as $\alpha_{\msbar}(q)$ with
$q=1/R$ since these schemes are so close to each other. Since we obtain
results consistent with the perturbative evolution, We can estimate the
continuum ratio $\sqrt K / \Lambda_{\msbar}$ = 1.72(13) for pure SU(3)
theory. Setting the scale using $\sqrt K$ = 0.44 GeV, then gives
$\Lambda_{\msbar}$=256(20) MeV.  These results are obtained for rather
modest energies ( $1/R \approx 1$--$3$ GeV ) but there is evidence from
studies in $SU(2)$ where higher energies have been reached~\cite{ukqcd}
that the method is stable as the energy scale is increased somewhat.
{}From lattice results for ratios of other non-perturbative quantities
(glueball masses, critical temperature, etc) to the string tension, one
can then determine their value in terms of $\Lambda_{\msbar}$ as well.

 Even though the scales probed in this work are relatively small
(i.e. only 3 GeV), the agreement with the perturbative evolution
of the coupling constant implies that this is a reasonable way to
determine the coupling constant in terms of non-perturbative
physical quantities. We are able to determine $\Lambda_{\msbar}$
relatively accurately compared to experiment. Of course experiment
has full QCD with dynamical light quarks included while precise
lattice  simulation of full QCD is still a considerable challenge.

\bigskip
\noindent{\bf Acknowledgements}
\par
The computations using the Meiko Computing Surface at Edinburgh
were supported by SERC under  grants
GR/G 32779 and GR/H 49191, by the University of Edinburgh and by
 Meiko Ltd. The computations on the Meiko Computing surface at
Liverpool were supported by SERC under grant GR/G 37132 and EC
contract SC1 *CT91-0642.
The computations using the CRAY were supported by SERC grant GR/H 01236.

\end{document}